\documentclass[%
 reprint,
nofootinbib,
 amsmath,amssymb,
aip,
]{revtex4-1}

\usepackage{graphicx}  
\usepackage{dcolumn}   
\usepackage{bm}        
\usepackage{siunitx}   
\usepackage{gensymb}   
\usepackage{booktabs}  

\begin{document}


\title{Highly efficient conversion of laser energy to hard X-rays in high intensity laser-solid simulations}
\author{S. Morris}
\email[]{sjm630@york.ac.uk}
\affiliation{York Plasma Institute, University of York, Heslington, YO10 5DQ, United Kingdom}
\author{A. Robinson}
\affiliation{Central Laser Facility, Rutherford-Appleton Laboratory, Chilton, Didcot OX11 0QX, United Kingdom}
\author{C. Ridgers}
\affiliation{York Plasma Institute, University of York, Heslington, YO10 5DQ, United Kingdom}

\date{\today}

\begin{abstract}

We present simulations which predict significantly higher laser to X-ray efficiencies than those previously found in high intensity ($10^{20}$-$10^{22}$ $\text{Wcm}^{-2}$) laser-solid simulations. The bremsstrahlung emission is shown to last for 10-100 ps, which is difficult to model with conventional particle-in-cell (PIC) codes. The importance of collective effects is also demonstrated, showing the limitations of Monte Carlo modelling in these systems. A new, open-source hybrid-PIC code with bremsstrahlung routines has been developed to model this X-ray production in 3D. Special boundary conditions are used to emulate complex electron refluxing behaviour, which has been characterised in 2D full-PIC simulations. The peak X-ray efficiency was recorded in thick gold targets, with 7.4\% conversion of laser energy into X-rays of energy 1 MeV or higher. The target size is shown to play a role in the conversion efficiency and angular distribution of emitted X-rays, and a simple analytic model is presented for estimating these efficiencies. 

\end{abstract}

\maketitle


\section{Introduction}\label{sec:intro}
When a high-intensity laser pulse strikes a solid target, the illuminated surface is ionised and forms a plasma layer. This plasma is further heated by the laser, injecting a large current of high energy (hot) electrons into the solid, with a roughly exponential energy distribution.\cite{intro:exponential_electron:Cowan} Multipetawatt laser facilities such as ELI \cite{laser:ELI} and Apollon \cite{laser:apollon} are expected to reach intensities between $10^{22}$-$10^{23}$ Wcm$^{-2}$, creating hot electrons over 100 MeV in energy. Such electrons could lead to efficient X-ray generation through either nonlinear Compton scatter (NCS) in the laser focus, or through bremsstrahlung as the electrons traverse the solid. These X-rays could act as a source for photonuclear reactions, \cite{motivation:photonuclear:Belyshev} radiotherapy, \cite{motivation:radiotherapy:Girolami} radiography \cite{motivation:radiography:Edwards} or in pair production for laboratory astrophysics. \cite{motivation:astro:Chen} As X-rays are emitted in the direction of motion for ultra-relativistic particles, angular distributions may also act as a diagnostic for electron motion and divergence within the solid.

Calculating the conversion efficiency of hot electron energy into bremsstrahlung radiation is complicated by the existence of competing energy loss mechanisms. Some energy goes to ionisation energy loss, where hot electrons excite atomic electrons in the solid and raise the target temperature. The hot electron current will also draw a resistive return current response, establishing fields which reduce the hot electron energy and further heat the target (Ohmic heating). Upon reaching the target edge, the highest energy electrons will escape, but the build-up of negative charge beyond the target boundary forms a sheath field which reflects most electrons back in. \cite{bremMC:Compant} Electron refluxing provides an additional energy loss mechanism, as the sheath field strength can change during a reflux.\cite{TNSA:rusby} These processes reduce the energy available for bremsstrahlung radiation, and must be characterised.

Several groups have already characterised the X-ray emission in laser-solid simulations by adding bremsstrahlung radiation to particle-in-cell (PIC) codes, and treating the solid as a cold, dense plasma.\cite{bremPIC:Sentoku, bremPIC:Ward, bremPIC:Pandit, bremPIC:Wan, bremPIC:Vyskocil, bremPIC:Wu, bremPIC:Martinez, bremPIC:Sawada} This approach has the advantage of directly modelling the absorption of laser energy in the pre-plasma, but requires a large number of computational macro-particles to suppress self-heating.\cite{bremPIC:Wu, code:epoch:Arber} In low resolution PIC simulations, electrons gain energy non-physically and bremsstrahlung routines allow this energy to be radiated away, producing a false X-ray emission. Due to high computational demands, these codes are typically restricted to short-pulse 2D simulations for thin targets, and are often not run long enough to capture the full bremsstrahlung emission. Previous attempts \cite{bremPIC:Wan, bremPIC:Vyskocil} to characterise the bremsstrahlung efficiency with PIC codes have considered the energy radiated in 36 fs and 300 fs, but these run-times are insufficient to capture an emission on the order 10-100 ps. 

Other groups \cite{bremMC:Henderson, bremMC:Sheng, bremMC:Armstrong, bremMC:Compant} have used Monte Carlo codes like \texttt{Geant4} \cite{code:geant4:2003, code:geant4:2006, code:geant4:2016} to model X-ray production in these systems. Electron injection characteristics are either modelled with PIC codes or assumed from the laser intensity, duration and focal spot size, and the bremsstrahlung emission is recorded as electrons propagate through the solid. Cross sections for bremsstrahlung, elastic scatter and ionisation energy loss are evaluated using the known values for electrons in solids, but each electron is treated independently and collective effects such as sheath-field energy loss, resistive electric fields and any generated magnetic fields are neglected.

We have developed a hybrid extension\cite{code:hybrid:url} to the PIC code \texttt{EPOCH},\cite{code:epoch:Arber,code:epoch:Ridgers,code:epoch:url} including resistive fields and elastic scatter equations,\cite{hybrid:Davies:1997,hybrid:Davies:2002} with additional bremsstrahlung and M{\"o}ller scatter algorithms adapted from \texttt{Geant4}. \cite{code:geant4:2003, code:geant4:2006, code:geant4:2016} This provides a similar functionality to the hybrid-PIC code \texttt{LSP}, \cite{code:LSP} but in an open-source format. A brief discussion of the code is presented in Section \ref{sec:code}, with technical details and benchmarking covered in the appendices. This code has allowed 3D simulation of the full bremsstrahlung emission with some collective effects, which cannot be done using traditional PIC or Monte Carlo codes. The bremsstrahlung radiation characteristics and boundary conditions are presented in Section \ref{sec:results}, where we show significantly higher laser-to-X-ray efficiencies than in previous simulations which only modelled run-times under 500 fs.\cite{bremPIC:Vyskocil, bremPIC:Wan} As the focus of this paper is energy loss within the solid, we will ignore the NCS X-rays associated with electron acceleration in the laser focal spot.\cite{NCS:Lezhnin, NCS:vyskovcil}

\section{Code}\label{sec:code}

\subsection{Hybrid-PIC} \label{subsec:hybrid_PIC}

Hybrid-PIC codes only simulate the hot electrons in laser-solid interactions, making them far less computationally expensive than traditional PIC codes.\cite{hybrid:Davies:1997,hybrid:Davies:2002} A hybrid field solver assumes the presence of a return current based on the temperature and resistivity of the solid, and adds in the corresponding fields without having to simulate the many cold particles in the solid density plasma. This cold plasma treatment is justified in metals and insulators, as target ionisation occurs quickly via pre-pulse heating, or field/collisional ionisation due to hot electrons.\cite{davies:ionisation} Equations describing the evolution of the fields, temperature and resistivity of the solid are presented in Appendix \ref{sec:code:solid}. 

Originally, hybrid-PIC codes were designed to track hot electrons of lower energy than those of interest here, lacking bremsstrahlung radiation and using a continuous form of ionisation energy loss.\cite{hybrid:Davies:2002} At higher electron energies, a more appropriate form of ionisation loss would also include discrete M{\"o}ller scatter, where incident electrons can lose large amounts of energy creating secondary hot electrons ($\delta$-rays) by fully exciting atomic electrons from the background solid. Algorithms for bremsstrahlung radiation have also been included. The full details of the physics applied to hot electrons as they traverse the solid are given in Appendix \ref{sec:code:hot_electrons}. Additionally, a set of benchmarks for the code is provided in Appendix \ref{sec:code:benchmarking}.

When hot electrons reach a simulation boundary, those above a critical energy $\kappa_\text{esc} a_0 m_e c^2$ are removed from the simulation (escaped), while the rest are reflected (refluxed). Here, $m_e$ and $c$ describe the electron rest mass and the speed of light respectively. The normalised vector potential 
\begin{align}
  a_0\approx 8.5\times 10^{-6} \sqrt{I_0\lambda^2} \label{eq:a0} 
\end{align}
describes the strength of the laser, with $I_0$ denoting the peak, cycle-averaged intensity ($\text{Wm}^{-2}$), and $\lambda$ the wavelength. Refluxing electrons have their total momentum reduced by $\kappa_\text{tnsa} a_0 m_e c$ on each re-injection, and are scattered through an angle randomly sampled from a uniform distribution between $\pm 0.5 \sigma_{\langle\Delta\theta\rangle}$. The empirical parameters are assigned values $\kappa_\text{esc}=2$, $\kappa_\text{tnsa}=2.7\times 10^{-3}$ and $\sigma_{\langle\Delta\theta\rangle}=23\degree$. These values were taken from 2D full-PIC simulations of electrons refluxing in sheath fields, and are defined in Section \ref{subsec:TNSA_result}.

\subsection{Simulation setup}

Hybrid-PIC simulations were run to model the hot electron to bremsstrahlung efficiency, $\eta_{e\rightarrow \gamma}$ for a variety of targets at different intensities. Unless stated otherwise, the hybrid-PIC simulations in this paper were run with cubic cells of side 0.7 \SI{}{\micro m}, as this was found to be the largest cell-size which converged the electric and magnetic fields in test runs. To improve statistics, the bremsstrahlung cross section was increased by a factor of 10, and macro-photon weights were reduced by the same factor to conserve real particle number. The efficiency of laser energy to hot-electron energy was set to $\eta_{l\rightarrow e}=0.3$. The initial background electron and ion temperatures were set to 300 K.

Hot electrons were injected into the simulation through the $x_\text{min}$ boundary, with spatial and temporal envelope functions, $f(r)$ and $g(t)$ respectively. A 2D Gaussian was used for $f(r)$, characterised by a radial full-width at half maximum (fwhm), $r_{fwhm}$. Similarly, a 1D Gaussian was used for $g(t)$, described by the fwhm, $t_{fwhm}$. To cut off low-weight macro-electrons, nothing was injected when $g(t)<0.1$, or into cells with $f(r)<0.5$. This gave a mean envelope of $\langle fg \rangle \approx 0.41$ for cells injecting electrons, with a mean root envelope $\langle \sqrt{fg} \rangle \approx 0.61$.

The laser intensity was varied in the range $10^{20}$-$10^{22}\text{ Wcm}^{-2}$, with $r_{fwhm} = 5$ \SI{}{\micro m}, and $t_{fwhm}=40$ fs. 1226 macro-electrons per timestep were injected into each cell which satisfied the envelope conditions. Macro-electrons were uniformly injected into a cone where the half angle was the smaller of $20\degree$ or $\tan^{-1}(\sqrt{2/(\gamma-1)})$ from Moore scaling,\cite{theory:Moore_scaling} where $\gamma$ is the Lorentz factor of a given injected electron.

\section{Results}\label{sec:results}

\subsection{Bremsstrahlung efficiency}\label{subsec:photon_result}

\begin{figure}
	\begin{center}
		\includegraphics[trim={0 0 0 0},clip,scale=0.6]{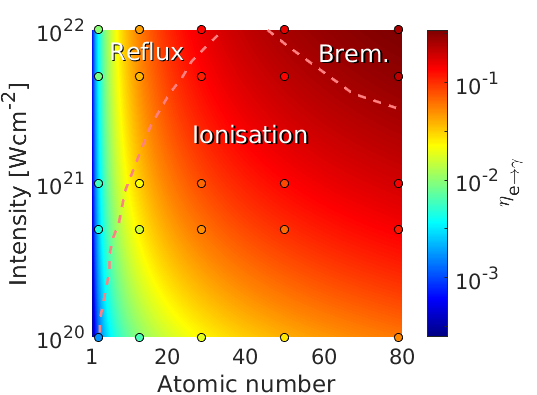}
	\end{center}
	\caption{Efficiency of hot electron kinetic energy to bremsstrahlung X-rays over 1 MeV photon energy in cubic targets of side-length $100$ \SI{}{\micro m}. The data-points show hybrid-PIC simulations, and the background heatmap comes from a simple scaling model. Regions where different energy loss mechanisms dominate are split by the pink lines.}
\label{fig:IZ_heatmap}
\end{figure}

FIG. \ref{fig:IZ_heatmap} shows $\eta_{e\rightarrow\gamma}$ evaluated for multiple $100\times 100\times 100$ \SI{}{\micro m^3} targets in 3D hybrid-PIC simulations. The target materials considered were Al, Cu, Sn and Au, and plastic CH targets are also shown plotted at atomic number $Z=2.7$. The peak efficiency of hot electron energy to X-rays over 1 MeV occurs for the $10^{22}$ \SI{}{Wcm^{-2}} shot on Au with $\eta_{e\rightarrow\gamma}=0.25$, which corresponds to a laser to X-ray efficiency of $\eta_{l\rightarrow\gamma}=0.074$. 

These simulations consider the full bremsstrahlung emission, and observe efficiencies higher than those reported from PIC simulations. Previous estimates\cite{bremPIC:Wan, bremPIC:Vyskocil} for $\eta_{l\rightarrow\gamma}$ in Al targets at $10^{22}$ \SI{}{Wcm^{-2}} have ranged from $4\times 10^{-6}$ to $8\times10^{-5}$ compared to 0.014 in these simulations, although the larger target size here also contributes to the greater efficiency. These high efficiencies are significant in experiments where bremsstrahlung is a background, suggesting measurement of X-rays from other processes (for example NCS) may be much more difficult than currently expected.

\begin{figure}
	\begin{center}
		\includegraphics[trim={0 0 0 0},clip,scale=0.5]{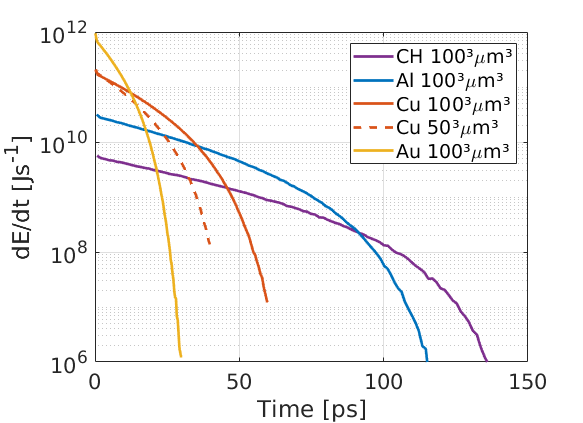}
	\end{center}
	\caption{Temporal distribution of bremsstrahlung radiation from hybrid-PIC simulations, with a laser of intensity $10^{22}$ \SI{}{Wcm^{-2}} on cubic targets of various compositions and sizes (labelled $l^3$ for side-length $l$).}
\label{fig:Xray:time}
\end{figure}

To estimate the run-times required to capture the full bremsstrahlung emission for FIG. \ref{fig:IZ_heatmap}, the X-ray characteristics were found for different targets shot by a $10^{22}$ \SI{}{Wcm^{-2}} pulse. FIG. \ref{fig:Xray:time} shows the rate of X-ray production for photons over 1 MeV in energy. The emission lasts on the order of 10-100 ps, with a strong dependence on target shape when using reflux boundaries, as electrons in smaller targets spend less time between reflux events and lose energy faster. The emission from lower $Z$ targets lasts longer, as ionisation loss and bremsstrahlung have lower stopping powers in these targets. This plot shows X-rays created within the solid and not the X-rays measured outside, as the code lacks target self-attenuation from the photoelectric effect (although this is less important for X-ray energies of a few MeV or greater). 

\begin{figure}
	\begin{center}
		\includegraphics[trim={0 0 0 0},clip,scale=0.35]{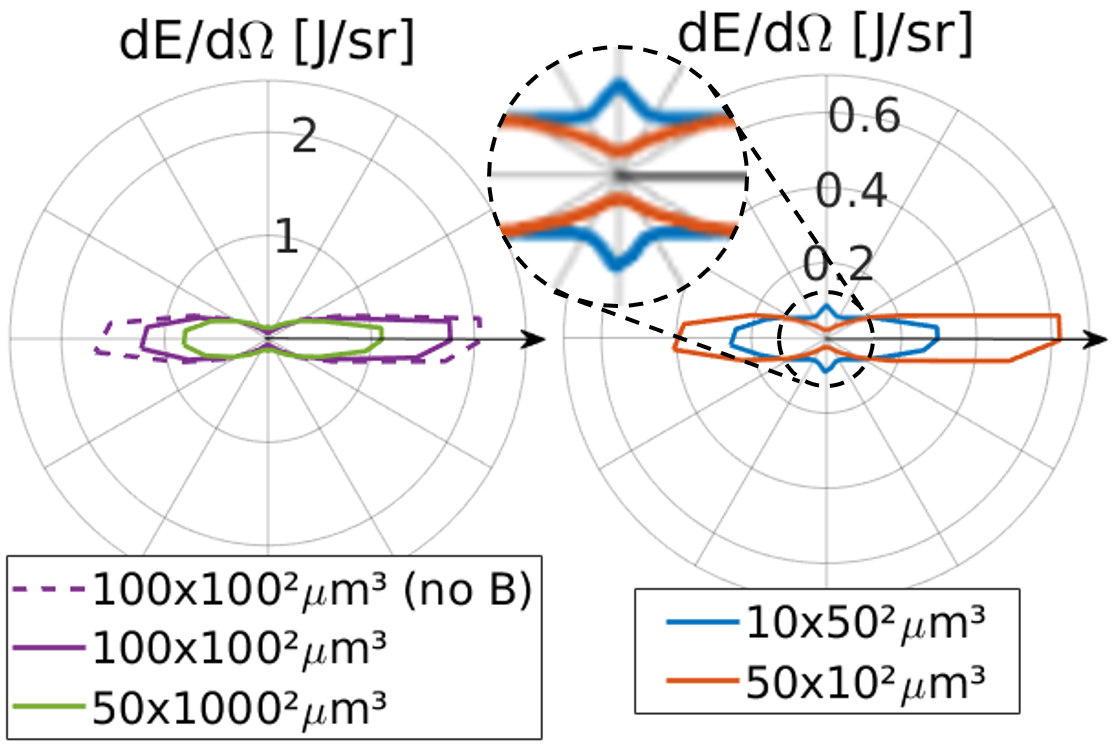}
	\end{center}
	\caption{Angular distribution of bremsstrahlung radiation from hybrid-PIC simulations ($10^{22}$ \SI{}{Wcm^{-2}}, Cu). The injection direction is given by the arrow, and the sign of $p_y$ determines the deviation direction for the macro-photons. Target dimensions are labelled as $l\times w^2$, where $l$ is the length parallel to electron injection, and the transverse area is $w\times w$. The dashed line data refers to a test where the magnetic field was held at 0 in all cells throughout the simulation.}
\label{fig:Xray:ang}
\end{figure}

The angular distribution of X-ray energy also varied with target geometry as shown in FIG. \ref{fig:Xray:ang}, although no target reproduced the lobes observed by Vysko\v{c}il \textit{et al}.\cite{bremPIC:Vyskocil} While the data does produce lobes when plotting energy per radian, $dE/d\theta$, in 3D simulations it is more appropriate to plot energy per steradian, $dE/d\Omega$ which re-weights the bins and shows a dominant emission in the forwards and backwards directions. A novel angular distribution is observed for the small foil $10\times 50^2$ \SI{}{\micro m^3}
target, which shows some emission perpendicular to the injection direction. This is because electrons deflected into the perpendicular direction can travel for a long time and emit many X-rays before hitting another boundary and scattering away. The perpendicular emission is less visible in the large foil 50 \SI{}{\micro m}$\times 1$ \SI{}{mm^2} target as electrons experience reflux scatter less often, and so more energy is lost by the time they scatter into a perpendicular direction. 

FIG. \ref{fig:Xray:ang} also shows that magnetic $B$ fields reduce the emission. While $B$ fields cannot take energy from the electrons, it was found that their presence led to more energy loss by resistive fields. This suggests $B$ fields reduce electron divergence, leading to higher current densities and stronger electric fields through (\ref{eq:Ohms_Law}).

\begin{figure}
	\begin{center}
		\includegraphics[trim={0 0 0 0},clip,scale=0.5]{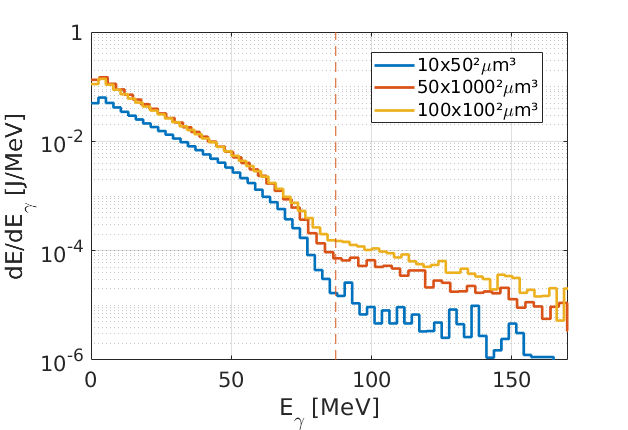}
	\end{center}
	\caption{Photon energy distribution of bremsstrahlung radiation from hybrid-PIC simulations ($10^{22}$ \SI{}{Wcm^{-2}}, Cu). Target dimensions are labelled as in FIG. \ref{fig:Xray:ang}. The pink line denotes the electron escape energy.}
\label{fig:Xray:energy}
\end{figure}

In FIG. \ref{fig:Xray:energy}, the bremsstrahlung energy spectra are given for some Cu targets. These spectra have a sharp gradient change at $E_\gamma \approx 86$ MeV, as the only electrons which can radiate above this energy escape the target after only one pass. The size of the target in $x$ determines the length of this pass, and the energy spectra beyond 86 MeV are grouped by this size. Smaller targets produce less bremsstrahlung radiation overall, as reflux events are more common and take away a greater proportion of the hot electron energy. 

The bremsstrahlung spectrum for the $10\times 50^2$ \SI{}{\micro m^3} Cu target was also calculated from a simulation without $\delta$-rays. Instead of adding $\delta$-rays as macro-electrons which can go on to produce photons, their energy was dumped to the local cell as a temperature increase. The resulting spectrum showed no significant difference to the case with $\delta$-rays, which suggests the rare high-energy photon emissions from rare high-energy $\delta$-rays play a negligible role in the total bremsstrahlung emission.  

To obtain these results, we have used special reflux boundaries which were characterised from full-PIC simulations which ran for 700 fs at the longest, as discussed in Section \ref{subsec:TNSA_result}. While escaping electrons leave the simulation in the first pass through the solid, it is unclear how well the refluxing data describes reflux events on time-scales where solid decompression becomes important. In Al $10^{22}$ \SI{}{Wcm^{-2}}, a final background temperature of 3.7 keV was recorded at the focal spot location, which suggests the target could grow 20 \SI{}{\micro m} over 100 ps according to the mean thermal ion speed.

\subsection{Energy loss mechanisms}\label{subsec:e_loss}

\begin{figure}
	\begin{center}
		\includegraphics[trim={0 0 0 0},clip,scale=0.35]{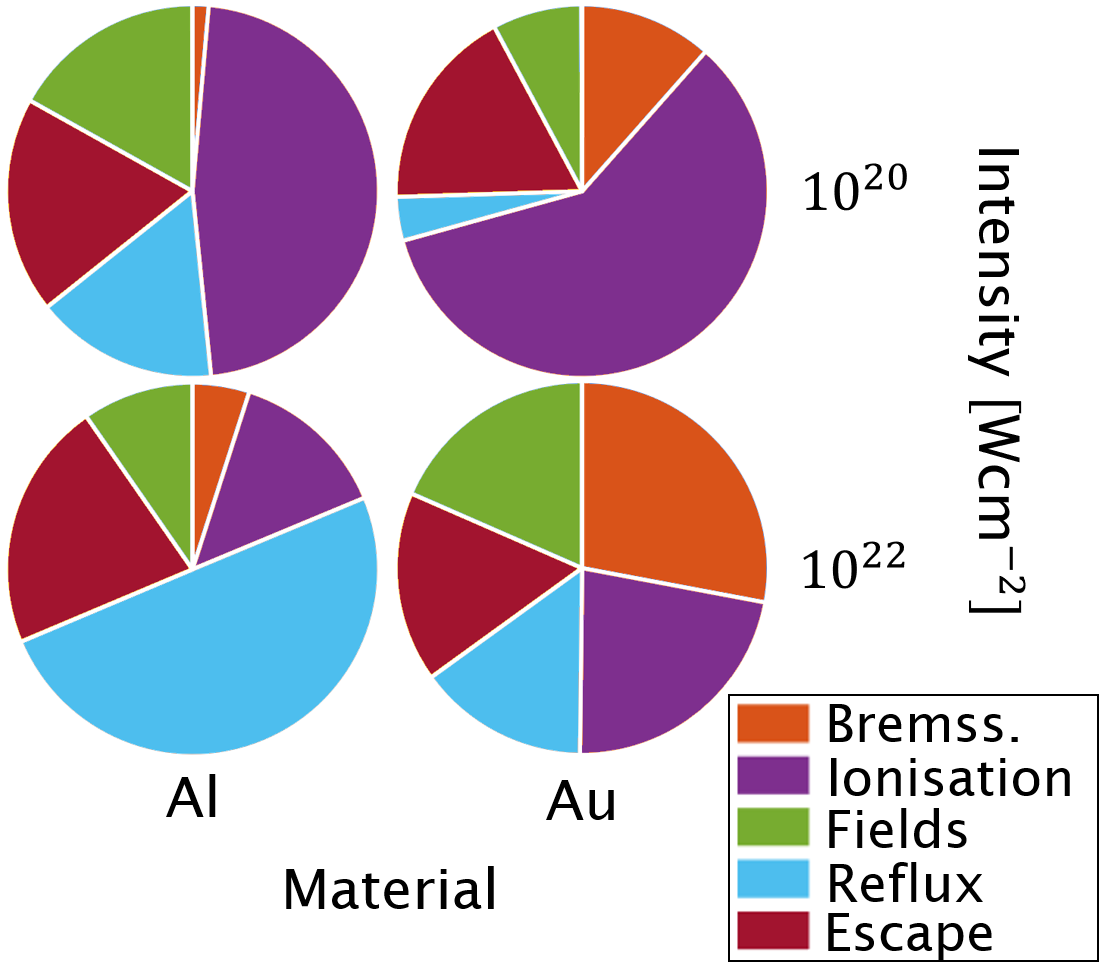}
	\end{center}
	\caption{Hot electron energy loss breakdown for four hybrid-PIC simulations from FIG. {\ref{fig:IZ_heatmap}}. The remaining electron energy in these simulations was less than $0.03\%$ of the total energy lost over the run-time.}
\label{fig:Energy_mechanisms}
\end{figure}

A breakdown of the total energy lost to each mechanism is shown in FIG. \ref{fig:Energy_mechanisms} for the $10^{20}$ and $10^{22}$ \SI{}{Wcm^{-2}} simulations in $100^3$ \SI{}{\micro m^3} Al and Au targets. It was found that 28 $\%$ of all lost energy in the Au $10^{22}$ \SI{}{Wcm^{-2}} simulation was due to bremsstrahlung radiation (from all photon energies), which dominated all other forms of energy loss. Ionisation loss dominated at $10^{20}$ \SI{}{Wcm^{-2}}, taking 47\% of the hot electron energy in Al and 59\% in Au. Reflux energy loss dominated in $10^{22}$ \SI{}{Wcm^{-2}} Al, accounting for 51\% of the energy loss. Escaping energy took away 17-22\% in all simulations, and resistive fields accounted for 8-19\%. While some electrons gained energy from these electric fields, field gains were less than 2\% of the field losses in all four simulations.  

A simple model was constructed to quickly estimate the efficiencies of hot electron energy loss mechanisms, and to demonstrate how these scale with laser and target parameters. This model condenses the exponential injection of electron kinetic energies, $\epsilon_k$ into three macro-electrons, characterised by the high-energy X-ray threshold, $\epsilon_\gamma^{th}$ (1 MeV here), and the escape energy $\epsilon_\text{esc}=\kappa_\text{esc} a_0 m_e c^2$. The ``warm" macro-electron describes all electrons with $\epsilon_k<\epsilon_\gamma^{th}$, the ``emitting" macro-electron holds $\epsilon_\gamma^{th} \leq \epsilon_k < \epsilon_\text{esc}$, and the ``escaping" macro-electron holds $\epsilon_k \geq \epsilon_\text{esc}$. The macro-electron weights are found from integrating
\begin{align}
  \frac{dN_e}{d\epsilon_k} = \frac{N_e}{\langle\epsilon_k\rangle} e^{-\epsilon_k/\langle\epsilon_k\rangle}
\end{align}  
between the defining kinetic energy limits, where the mean injected kinetic energy $\langle\epsilon_k\rangle = a_0 m_e c^2 \langle\sqrt{fg}\rangle$. From (\ref{eq:N_cell}), the total number of injected electrons is
\begin{align}
   N_e = \frac{I_0 \langle fg\rangle \left(\frac{\pi  r_{fwhm}^2}{4}\right) \left(t_{fwhm}\sqrt{\frac{\ln10}{\ln2}}\right) \eta_{l\rightarrow e}}{\langle\epsilon_k\rangle}
\end{align}
after substitution of the full injection area and pulse duration for our envelopes. Similarly, the three macro-electron $\epsilon_k$ values are found from integrating $\epsilon_k dN_e/d\epsilon_k$ between the defining energies. 

Four stopping-powers are used to characterise energy loss from the individual energy loss mechanisms. For bremsstrahlung \cite{theory:Jackson} and ionisation energy loss,\cite{hybrid:Davies:2002} we use the continuous stopping power approximations,
\begin{align}
  -\left.\frac{d\epsilon}{dx}\right|_{brem} =& \frac{\gamma e^6}{12\pi^3\epsilon_0^3 m_e c^3 \hbar} n_i Z^2 \ln\left(\frac{5.6 \pi \epsilon_0\hbar c}{Z^{1/3}e^2}\right) \label{eq:stop_brem}
  \\ 
  -\left.\frac{d\epsilon}{dx}\right|_{ion} =& \frac{Zn_i e^4}{4\pi\epsilon_0^2 m_e v^2}\left(\ln\left(\frac{\epsilon_k}{I_{ex}}\right) + \frac{1}{2}\ln(\gamma + 1) \right.
  \notag\\ \label{eq:stop_ion}
  &+ \left.\frac{0.909}{\gamma^2} - \frac{0.818}{\gamma} - 0.246\right)
\end{align}
where the bremsstrahlung stopping power considers emission into photons of all energies. These use the permittivity of free space, $\epsilon_0$, reduced Planck constant, $\hbar$, and electron charge and speed, $e$ and $v$ respectively. Solid parameters $n_i$ and $I_{ex}$ denote the ion number density and mean excitation energy. The stopping power from  photons over energy $\epsilon_\gamma^{th}$ is
\begin{align}
 -\left.\frac{d\epsilon}{dx}\right|_{\epsilon_\gamma > \epsilon_\gamma^{th}} =-\left.\frac{d\epsilon}{dx}\right|_{brem}(\epsilon_k - \epsilon_\gamma^{th}) 
\end{align}
which can be used to calculate $\eta_{e\rightarrow\gamma}$. In a target of size $L_x \times L_y \times L_z$, the typical path between two boundaries is roughly $\frac{1}{3}(L_x + L_y + L_z)$. The energy lost in a reflux event is described by the reflux boundaries, so we approximate a continuous reflux stopping power of the form
\begin{align}
  \label{eq:stop_tnsa}
  -\left.\frac{d\epsilon}{dx}\right|_{tnsa} = \frac{3\kappa_\text{tnsa} a_0 m_e c^2}{L_x + L_y + L_z}.
\end{align}
For fields, the stopping power is equivalent to the Lorentz force $-eE$, where the electric fields in this system are described in (\ref{eq:Ohms_Law}). Assuming the hot electron current density, $j_h$ is balanced by the background electron current, the stopping power may be written as $-e\eta j_h$ for a solid with resistivity, $\eta$. By approximating a suitable form for $j_h(x)$, we have
\begin{align}
   \label{eq:stop_field}  
   -\left.\frac{d\epsilon}{dx}\right|_{field} = e^2 \langle\eta\rangle \frac{I_0 \langle fg \rangle \left(\frac{1}{2}r_{fwhm}\right)^2\eta_{l\rightarrow e}}{(x \tan\theta_c + \frac{1}{2}r_{fwhm})^2\langle\epsilon_k\rangle }
\end{align}
where a constant typical resistivity $\langle\eta\rangle$ has been used. Here we have assumed the injected current begins with a circular area of radius $r_{fwhm}/2$, where electrons move into a cone of half-angle $\theta_c$, such that the current radius at $x$ includes the $x \tan\theta_c$ term. This ensures the field stopping power diminishes as electrons spread out in the solid.

The ``warm" and ``emitting" macro-electrons are integrated through these stopping powers until they have no more energy, and the ``escaping" macro-electron is integrated to $x=L_x$, at which point the remaining energy is considered to be escaped. Using $\theta_c = 20\degree$ and $\langle\eta\rangle=10^{-6}$ \SI{}{\Omega m}, $\eta_{e\rightarrow\gamma}$ was calculated over the simulation range shown in FIG. \ref{fig:IZ_heatmap}, and forms the background heatmap. This simple model shows good agreement with the simulation data. 

While calculating this heatmap, a constant $n_i=6\times10^{28}$ \SI{}{m^{-3}} was used, along with the approximation $I_{ex}\approx 11eZ$. The dominant emission mechanisms were identified in each calculation, and are grouped by the pink lines in FIG. \ref{fig:IZ_heatmap}. For high $Z$ targets, ionisation loss dominates at low intensities, while bremsstrahlung dominates at high intensities. In lower $Z$ targets, the stopping power associated with these processes decreases and reflux energy loss becomes the dominant process, making these set-ups especially unsuitable for modelling using traditional Monte Carlo codes which lack collective effects.

\subsection{Reflux energy loss}\label{subsec:TNSA_result}

The reflux boundaries described in Section \ref{subsec:hybrid_PIC} are characterised by three empirical parameters. These are related to the escape energy threshold, $\kappa_\text{esc}$, the mean reflux momentum loss, $\kappa_\text{tnsa}$ and the range of scatter values, $\sigma_{\langle\Delta\theta\rangle}$. The specific values  used in the 3D hybrid-PIC simulations of sections \ref{subsec:photon_result} and \ref{subsec:e_loss} were calculated from 2D full-PIC simulations in \texttt{EPOCH}. These full-PIC simulations model electron refluxing in the sheath fields, and are similar to those performed by Rusby \textit{et al}. \cite{TNSA:rusby}

Four simulations modelled two targets (C and Au) shot at two different laser intensities ($10^{20}$ and $10^{22}$ \SI{}{Wcm^{-2}}). All targets were given a pre-plasma for $(-4<x<0)$ \SI{}{\micro m}, with an electron number density $n_e(x) = n_{e0} \exp(x/L_p)$, where $n_{e0}$ is the solid electron number density, and the pre-plasma scale-length, $L_p =2$ \SI{}{\micro m}. The solid density region spanned $0<x<L_s$, where the solid length, $L_s$ was 10 \SI{}{\micro m} for C and 2 \SI{}{\micro m} for Au. All simulations used a laser with $r_{fwhm}=$ 5 \SI{}{\micro m}, and $t_{fwhm}=40$ \SI{}{fs} to match the hybrid-PIC simulations.

C simulations assumed fully ionised targets, with a run-time of 700 fs, 250 macro-electrons and 50 macro-ions per cell, for square cells of side 20 nm. The simulation window spanned $(-30 < x < 130)$ \SI{}{\micro m}, with $y$ between $\pm10$ \SI{}{\micro m}. For Au we used $\text{Au}^{51+}$ ions, which had a greater $n_e$ than Al and so smaller square cells of side 5 nm were used to prevent self-heating. The simulation window was reduced to having $x$ and $y$ range between $\pm10$ \SI{}{\micro m} and $\pm4$ \SI{}{\micro m} respectively, with 125 macro-electrons and 25 macro-ions per cell, and a run-time of 160 fs.

Six \texttt{EPOCH} particle probes were positioned in the simulation window: two tracking electrons escaping the window through $x_\text{min}$ and $x_\text{max}$ boundaries, and the rest tracking particles entering and leaving the solid density region at $x=0$ and $x=L_s$. These probes output the momentum, position and weight of each macro-particle passing them, and have been extended to also output particle ID and crossing time. All macro-electrons in the pre-plasma were tracked, and were considered hot electrons once they triggered the $x=L_s$ probe for the first time.

Once hot electrons exit the solid-density region there are four possible end-states: re-entering the solid (refluxed), escaping the simulation window through an $x$ boundary (escaped), a $y$ boundary (lost, no useful information), or remaining outside the solid but within the window until the end of the simulation (absorbed into the sheath field). These runs determined the likelihood of these end-states, and also looked at how the properties of hot electrons changed over a reflux.

\begin{figure}
	\begin{center}
		\includegraphics[trim={0 0 0 0},clip,scale=0.5]{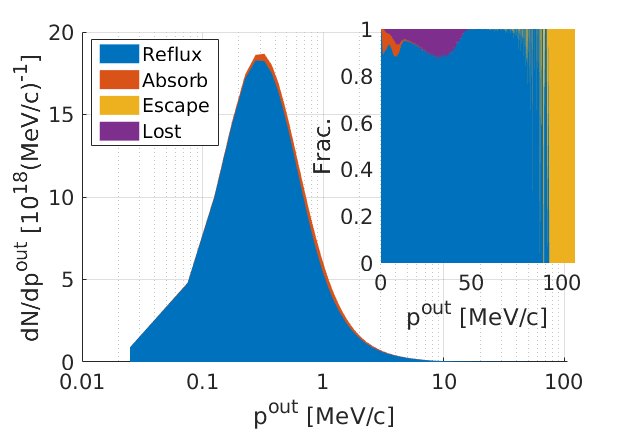}
	\end{center}
	\caption{The number spectrum of hot electrons exiting the solid through $x=L_s$ in the C $10^{22}$ \SI{}{Wcm^{-2}} simulation, binned by the electron momentum. The colour denotes the distribution of end-states in each bin. An insert is provided to show the fate of rare, high momentum outgoing electrons.}
\label{fig:TNSA:fate}
\end{figure}

FIG. \ref{fig:TNSA:fate} shows a number spectrum of hot electrons in the C $10^{22}$ \SI{}{Wcm^{-2}} simulation as they pass $x=L_s$ into the vacuum. Most electrons reflux back into the target, with the highest energy electrons escaping through $x_\text{max}$ and some lower energy electrons ending in the sheath field. The $\kappa_\text{esc}$ parameter was chosen such that $\kappa_\text{esc}a_0 m_e c^2$ was the energy associated with the first bin in FIG. \ref{fig:TNSA:fate} which had all electrons escape after passing into the vacuum from $x=L_s$. The sharp switch of end-states justifies our treatment of a critical escape energy. Electrons labelled as lost have escaped through a $y$ boundary, and it is unclear whether they would have refluxed, escaped, or been absorbed if the simulation window was larger in $y$.

The simulation was repeated with a smaller window of size $(-30<x<90)$ \SI{}{\micro m}, and hot electrons were found to escape at the same energy as in the larger window simulations, with similar energy distributions at $x_\text{max}$. This suggests convergence in the escape energy cut-off, but these 2D sheath fields will decay slower with distance than fields in 3D space, so this cut-off is likely an over-estimate. The qualitative behaviour is similar in all four simulations for electrons exiting through $x=L_s$. Refluxing also dominates electrons exiting the solid on the pre-plasma side through $x=0$, but the absorption chance is typically greater on this boundary. For example, C $10^{22}$ \SI{}{Wcm^{-2}} has 20\% absorption in the bin corresponding to the $dN/dp$ peak for the $x=0$ probe, compared to 2\% for the peak bin in the $x=L_s$ probe. A typical reflux event was found to last $\sim$60 fs on the pre-plasma side, but only $\sim$20 fs on the rear, so this increased absorption could be due to counting more refluxing electrons outside the solid at the simulation end.

\begin{figure}
	\begin{center}
		\includegraphics[trim={0 0 0 0},clip,scale=0.4]{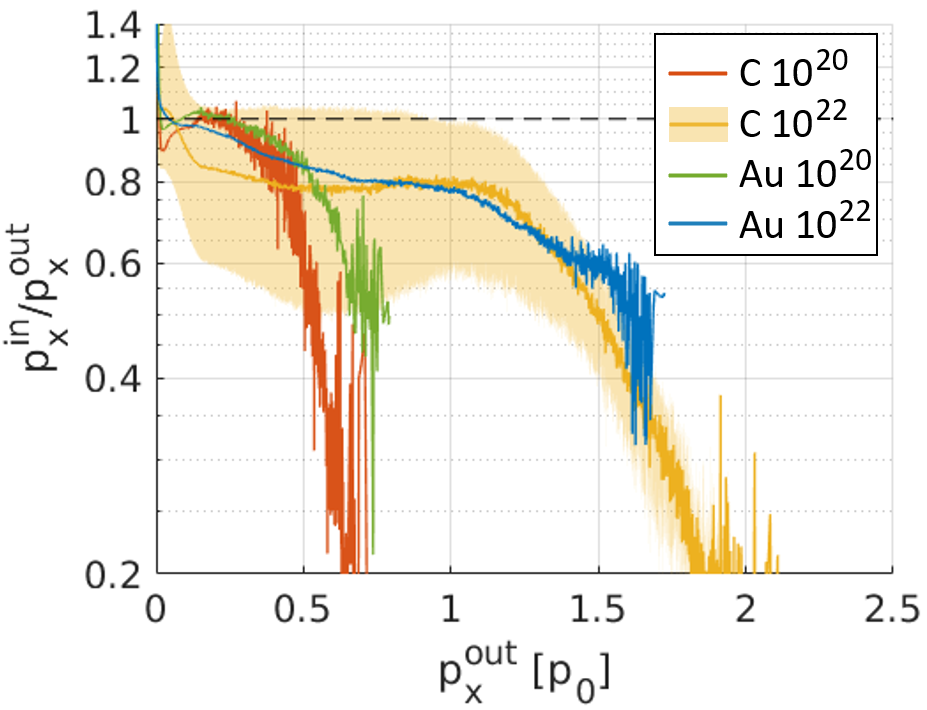}
	\end{center}
	\caption{Refluxing electrons on $x=L_s$ are binned by outgoing longitudinal momentum (in units of the ponderomotive momentum $p_0=a_0m_ec$), and the bin averaged momentum change is plotted. Simulations are labelled by target material and laser intensity in \SI{}{Wcm^{-2}}. The shaded region on C $10^22$ denotes the upper and lower average deviations from the mean in each bin.}
\label{fig:TNSA:loss}
\end{figure}

The exiting, $p_x^\text{out}$ and returning, $p_x^{in}$ longitudinal momenta of electrons leaving and re-entering the solid through $x=L_s$ was recorded for all reflux events. In FIG. \ref{fig:TNSA:loss}, refluxing electrons are binned by $p_x^{out}$ values, and the mean fractional change of longitudinal momentum in each bin is shown by the solid lines for all four simulations. Most hot electrons lose longitudinal momentum when refluxing, with the highest energy electrons losing the most. The $\kappa_\text{tnsa}$ parameter is chosen such that $\kappa_\text{tnsa}a_0 m_e c$ is the average momentum loss for all hot electrons exiting and re-entering the solid, on both the $x=0$ and $x=L_s$ sides. Electron momenta beyond those plotted in this figure mostly escape the simulation window through the $x_{max}$ boundary. These trends seem similar across the different target materials, sizes and run-times, although the results appear grouped by intensity.

In C simulations, it was found that $\eta_{l\rightarrow e}$ was 0.27 in the $10^{22}$ \SI{}{Wcm^{-2}} run, but only 0.03 for $10^{20}$ \SI{}{Wcm^{-2}}. This demonstrates different injection characteristics for FIG. \ref{fig:TNSA:loss}, and could explain the reduced peak momentum achieved in lower intensity runs (relative to the ponderomotive momentum). The $\eta_{l\rightarrow e}$ value for $10^{20}$ \SI{}{Wcm^{-2}} is similar to the one found to fit the data for the FIG. \ref{fig:benchmark:Evans} benchmark in Appendix \ref{sec:code:benchmarking}, which was at $3.1\times10^{20}$ \SI{}{Wcm^{-2}} and also at normal incidence. At oblique incidence, the FIG. \ref{fig:benchmark:Clarke} benchmark fit with $\eta_{l\rightarrow e}=0.3$, which is closer to that of C $10^{22}$ \SI{}{Wcm^{-2}}, despite only being at $4\times 10^{20}$ \SI{}{Wcm^{-2}}. The choice to set $\eta_{l\rightarrow e}=0.3$ in sections \ref{subsec:photon_result} and \ref{subsec:e_loss} was made to allow for direct comparisons between the results.

\begin{figure}
	\begin{center}
		\includegraphics[trim={0 0 0 0},clip,scale=0.38]{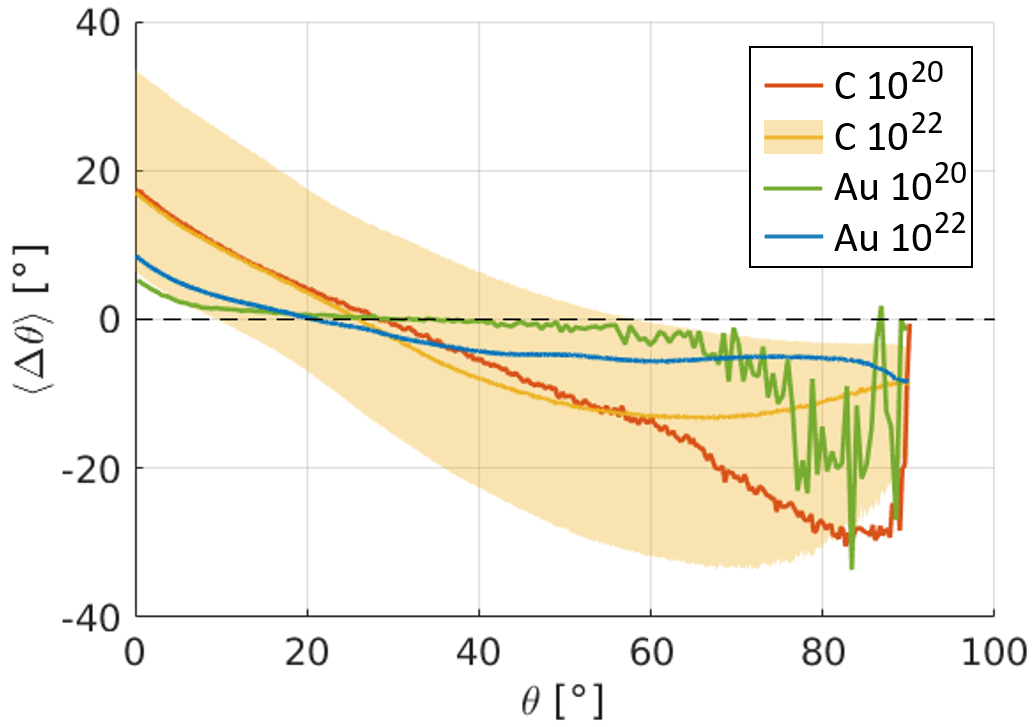}
	\end{center}
	\caption{Refluxing electrons at $x=L_s$ are binned by outgoing $\theta^\text{out}=\tan^{-1}(|p_y^\text{out}/p_x^{out}|)$, and the $\Delta\theta = \theta^\text{in}-\theta^\text{out}$ values are averaged in each bin. Simulations are labelled as in FIG. \ref{fig:TNSA:loss}. The shaded region for C $10^{22}$ denotes the upper and lower average deviations from $\langle\Delta\theta\rangle$ in each bin.}
\label{fig:TNSA:ang}
\end{figure}

In addition to the large decrease in $p_x$, a smaller increase in $p_y$ was found on refluxing which contributes to the angle increase observed by Vysko\v{c}il \textit{et al}. \cite{bremPIC:Vyskocil} FIG. \ref{fig:TNSA:ang} shows the average change in angle when refluxing for hot electrons exiting the solid at different angles. On average, hot electrons exiting below $30\degree$ to the injection direction return at a greater angle, and those above $30\degree$ return lower. The shaded-region which denotes average deviation from the mean is large and roughly uniform over all outgoing angles, which shows a large range of scatter angles independent of the outgoing direction. The $\sigma_{\langle\Delta\theta\rangle}$ parameter is the shaded-region size for $\langle\Delta\theta\rangle$, averaged over all bins for electron reflux events at both $x=0$ and $x=L_s$. This average is weighted by the number of electrons in each bin.

The empirical parameters have been calculated in each simulation, and the results are shown in TABLE \ref{table:tnsa_param}. Typical averages were chosen for $\kappa_\text{tnsa}$ and $\sigma_{\langle\Delta\theta\rangle}$ for our hybrid-PIC simulations. We also chose $\kappa_\text{esc}=2$ to be closer to the C $10^{22}$ \SI{}{Wcm^{-2}} simulation, as this has the most similar $\eta_{l\rightarrow e}$ value to our hybrid electron injection.

\begin{table}[h!]
  \centering
  \begin{tabular*}{0.35\textwidth}{@{\extracolsep{\fill}} l c c c}
    \\ [-1ex]
    \hline \\ [-2ex]
    \multicolumn{1}{c}{Run} & \multicolumn{3}{c}{Parameters} \\
    & $\kappa_\text{esc}$ & $\kappa_\text{tnsa}$ & $\sigma_{\langle\Delta\theta\rangle}$ [$\degree$] \\
    \hline \\ [-1.5ex]
      $10^{20}$, C  & 0.81  & $1.7\times 10^{-3}$ & 27 \\
      $10^{20}$, Au & 0.75 & $1.6\times 10^{-3}$ & 10 \\
      $10^{22}$, C  & 1.9 & $4.2\times 10^{-3}$ & 32 \\
      $10^{22}$, Au & 1.6 & $3.2\times 10^{-3}$ & 22 \\ [+0.75ex]
    \hline
  \end{tabular*}
  \caption{Reflux boundary characterisation parameters from 2D full-PIC simulations, labelled by laser intensity in \SI{}{Wcm^{-2}}, and target material.}
  \label{table:tnsa_param} 
\end{table}

\section{Conclusion}

A hybrid-PIC code has been written and benchmarked against experiments for Vulcan shots around $10^{20}$ \SI{}{Wcm^{-2}}, where hot electron injection was found to form the dominant source of uncertainty. Full-PIC simulations in 2D demonstrated that over short time-scales (up to 700 fs from a 40 fs pulse) most electrons reflux with an energy loss, and the highest energy electrons escape. 

Using our reflux boundaries, the bremsstrahlung emission occurred over a time-scale on the order of 10-100 ps, and showed higher efficiencies than previously reported. PIC simulations were shown to underestimate the bremsstrahlung efficiency by orders of magnitude, as they are unable to capture the full emission. Monte Carlo codes are expected to overestimate the emission, as they lack collective energy loss mechanisms. In these 3D simulations, we did not observe the lobes in the bremsstrahlung angular distribution found in 2D full-PIC simulations.

Different energy-loss mechanisms were found to dominate at different laser intensities and target atomic numbers, with bremsstrahlung dominating in high-intensity high-$Z$ set-ups. A simple analytic model was provided for estimating efficiencies $\eta_{e\rightarrow\gamma}$, and showed good agreement with the predictions of the code. 

At these time-scales, the code could be improved by evolving the immobile background ion fluid with a hydrodynamic code, and diffusing the solid temperature with a thermal conductivity model. 2D full-PIC simulations could be set-up starting with hot electrons in decompressed targets, to better characterise refluxing at later times. The code could also be extended to include photon transport effects like photoelectric attenuation, and Bethe-Heitler pair production \cite{theory:Bethe_Heitler} to better model higher intensities. 

\textbf{Acknowledgements}: This work was in part funded by the UK EPSRC grants EP/G054950/1,
EP/G056803/1, EP/G055165/1, EP/ M022463/1
 and EP/M018156/1.
 
The project was undertaken on the Viking Cluster, which is a high performance compute facility provided by the University of York. We are grateful for computational support from the University of York High Performance Computing service, Viking and the Research Computing team.

We would also like to thank our AWE partners N. Sircombe (now at Arm Ltd.) and G. Crow and for their support in this project.

The data that support the findings of this study (code, input decks and additional benchmarks) are openly available in the York Research Database\cite{data_availability} at http://doi.org/10.15124/707baa95-44e0-4f55-9476-ef1097b0a668.

\bibliographystyle{unsrt}
\bibliography{paper_bib.bib}

\appendix

\section{Solids}\label{sec:code:solid}
Hybrid-PIC codes model the transport of hot electrons through solids with a significantly colder and denser electron population. The hot electron currents in these systems exceed the Alfv{\'e}n limit, and propagate by drawing a return current density, $\textbf{j}_b$ from the background electrons.\cite{intro:returnCur:Hammer} This return current establishes a resistive electric field, $\textbf{E}$ according to Ohm's law
\begin{align} \label{eq:Ohms_Law}
  \textbf{E} = \eta \textbf{j}_b,
\end{align}
where $\eta$ denotes the local resistivity of the solid. To avoid simulation of background particles, the field equations are expressed using only the hot electron current density, $\textbf{j}_h$ by substituting the total current density $\textbf{j} = \textbf{j}_h + \textbf{j}_b$ into the Amp{\`e}re-Maxwell law, and iterating the magnetic field $\textbf{B}$ with the Faraday-Lenz law
\begin{align}
  \textbf{E} &= \eta \left( \frac{1}{\mu_0}\nabla \times \textbf{B} - \textbf{j}_h \right) 
  \label{eq:Ampere_Maxwell}  
  \\
  \frac{\partial\textbf{B}}{\partial t} &= - \nabla \times \textbf{E} .
  \label{eq:Faraday_Lens}
\end{align}
The displacement current in (\ref{eq:Ampere_Maxwell}) has been negelcted, as this is negligible over multi-picosecond timescales. \cite{hybrid:Davies:1997}

Our code was built as an extension to \texttt{EPOCH} by introducing a new solid concept to the code. Solids are single-element immobile fluids added to the simulation window, and are described by an atomic number, $Z$, mean excitation energy, $I_{ex}$, and radiation length, $X_0$. A spatially varying ion number density, $n_i$ is assigned to each solid, and multiple solids may be assigned to the same cell to construct compound materials like plastic.

The hybrid mode also tracks the local background electron and ion temperatures, $T_e$ and $T_i$ (in Kelvin) and the resistivity in each cell. The temperature-dependent resistivity is calculated using a reduced form of the Lee-More model, \cite{eta:LeeMore}
\begin{align} \label{eq:LeeMoreGeneral}
  \eta = \frac{m_e}{Z^* n_i e^2 \tau A^\alpha}
\end{align}
where $Z^*$ is the local solid ionisation state (given by the More Table IV algorithm,\cite{eta:More:TableIV}) $\tau$ is the electron relaxation time, and $A^\alpha$ is a correction factor. Here the Lee-More equations have been converted to SI units. Our reduced model varies between the hot and cold relaxation time limits
\begin{align}
  (\tau A^\alpha)_{\text{hot}} &= \frac{128\pi\epsilon_0^2}{3e^4}\sqrt{\frac{m_e}{2\pi}}\frac{(k_bT_e)^{3/2}}{{(Z^*)}^2n_i\ln\Lambda} \\
  (\tau A^\alpha)_{\text{cold}} &= \frac{R_0}{\bar{v}}\lambda_1
\end{align}
where the Coulomb logarithm $\ln\Lambda$ is evaluated using the Lee-More method, \cite{eta:LeeMore} the ion sphere radius $R_0=(3/4\pi n_i)^{1/3}$, mean thermal speed $\bar{v}=\sqrt{3 k_b T_e/m_e}$, and $\lambda_1$ is a fitting parameter. The value of $\tau A^\alpha$ used in (\ref{eq:LeeMoreGeneral}) is $\max{((\tau A^\alpha)_{\text{hot}},(\tau A^\alpha)_{\text{cold}})}$, and resistivity is taken to be $\eta\lambda_2$, where $\lambda_2$ is a second fit parameter. The $(\lambda_1, \lambda_2)$ values are taken to be (7, 3.5) from a fit to experimental Al resistivities. \cite{eta:Al:Milchberg} 

The electron temperatures of the background solid are updated for each cell and timestep according to
\begin{align} \label{eq:Te_rise}
  \Delta T_e = \frac{\rho_\epsilon}{Zn_iCk_b}
\end{align}
where $\rho_\epsilon$ is the density of the energy deposited in the cell over the timestep, $\Delta t$, and $C$ is the heat capacity of the solid
\begin{align}
  C = 0.3 + 1.2T'\frac{2.2 + T'}{(1.1 + T')^2}
\end{align}
for $T'=(k_bT_e/e)Z^{-4/3}$. \cite{theory:heat_capacity} For compound solids, we replace the electron number density term $n_e = Z n_i$ in (\ref{eq:Te_rise}) with the sum of $n_e$ over all solids in the cell, and calculate a cell-averaged $1/C$ value weighted by the $n_e$ value of each solid. This ensures that two overlapping solids of the same material retains the same behaviour as the equivalent single solid.

In Ohmic heating, the induced return current dissipates heat by travelling through the resistive solid, depositing an energy density of $\rho_\epsilon = \eta \textbf{j}_h \cdot \textbf{j}_h\Delta t$ (as $|\textbf{j}_h|\approx|\textbf{j}_b|$). \cite{hybrid:Davies:2002} In ionisation heating, $\rho_\epsilon$ is the sum of the ionisation losses for all hot electrons in a cell over $\Delta t$, divided by the cell volume.

Background electrons share energy with the ions through collisions, updating the temperatures of each species at the rates
\begin{align}
  \frac{dT_e}{dt} = (T_i-T_e)\frac{(Z^*)^2 e^4 n_i}{t_c}
  \\
  \frac{dT_i}{dt} = (T_e-T_i)\frac{(Z^*)^3 e^4 n_i}{t_c}
\end{align}
with the repeated term, $t_c$ representing
\begin{align}
  \frac{1}{t_c} = \frac{2}{3(2\pi k_b)^{3/2}} \frac{\sqrt{m_e m_i}\ln\Lambda}{\epsilon_0^2(T_e m_i + T_i m_e)^{3/2}}
\end{align}
where $T_i$ and $m_i$ describe the ion temperature and mass respectively. \cite{theory:equilibration} 

\section{Hot electrons}\label{sec:code:hot_electrons}

Hot electrons are injected into the simulation with exponentially distributed energies, and a mean kinetic energy given by ponderomotive scaling, $\langle\epsilon_k(\textbf{r},t)\rangle = a(\textbf{r},t)m_ec^2$, for position $\textbf{r}$ and time $t$. Here we use a local normalised vector potential $a(\textbf{r},t)=a_0 \sqrt{f(\textbf{r})g(t)}$, which applies an intensity reduction to (\ref{eq:a0}) due to the envelope functions $f(\textbf{r})$ and $g(t)$. The number of electrons injected into the simulation, $N_e^\text{cell}$ from a cell with transverse area, A, over a time-step, $\Delta t$ is given by
\begin{align} \label{eq:N_cell}
  N_e^\text{cell} = \frac{I_0f(\textbf{r})g(t)A\Delta t \eta_{l\rightarrow e}}{\langle\epsilon_k(\textbf{r},t)\rangle}
\end{align}
where $\eta_{l\rightarrow e}$ is the absorption efficiency of laser energy into hot electron kinetic energy.

The ionisation energy loss algorithm has been adapted from \texttt{Geant4}, and takes two forms depending on the energy transferred to background electrons. \cite{code:geant4:2003, code:geant4:2006, code:geant4:2016} Hot electron energy loss is described by a continuous stopping power when background electrons are excited to energies, $\epsilon_k^\delta$ less than a cut-off energy, $\epsilon_{k,\text{cut}}$,
\begin{align}
  \frac{dE}{dx}
  = 
  \frac{Z n_i e^4}{8 \pi \epsilon_0^2 m_e v^2}
  \left(
    \ln\left(\frac{2(\gamma+1)m_e^2c^4}{I_{ex}^2}\right) + F^- - \delta)
  \right)
\end{align}
where $v$ and $\gamma$ are the speed and Lorentz factor of the hot electron respectively, and $\epsilon_{k,\text{cut}}$ is set to 1 keV. Here, $F^-$ is a function of $\gamma$ and $\epsilon_{k,\text{cut}}$, and $\delta$ is the density effect function. \cite{theory:Moller_scatter} Background electrons excited to energies over $\epsilon_{k,\text{cut}}$ are treated as a discrete emission ($\delta$-rays), and are added into the simulation as macro-electrons.

Secondary particle emission from macro-electrons in a PIC code is achieved using the optical depth method. \cite{theory:PIC_MC_emission} Over timestep $dt$ in a solid with a cross section per atom $\sigma$, a macro-electron covers an optical depth, $d\tau = n_i\sigma v dt$, where $d\tau$ is equivalent to the probability of an emission event during $dt$. The cumulative probability of emission by optical depth $\tau$ is $F(\tau)=1-e^{-\tau}$. Hence, an optical depth of emission, $\tau_e$ can be sampled for each macro-electron using $\tau_e=-\ln(1-x_r)$, where $x_r$ is a uniformly-distributed random number between 0 and 1. The total $\tau$ traversed by a macro-electron is saved, and once this exceeds $\tau_e$ a secondary particle is emitted, the saved $\tau$ value is reset and a new $\tau_e$ is sampled.

Discrete $\delta$-ray emission uses the cross section of high energy M\"{o}ller scatter, \cite{theory:Moller_scatter} and a \texttt{Geant4} algorithm for sampling the $\delta$-ray energy from the differential cross section. \cite{code:geant4:2003, code:geant4:2006, code:geant4:2016} A separate optical depth is used for tracking bremsstrahlung photon emission, which is characterised using the Seltzer-Berger differential cross-sections. \cite{theory:Seltzer_Berger} Following the theory of Wu \textit{et al}, \cite{bremPIC:Wu} the Seltzer-Berger cross sections are enhanced by the factor, $F_\sigma$
\begin{align}
F_\sigma = 1 + \frac{\ln\left|\lambda_D/a_s\right|}{\ln |a_s m_e c^2/\hbar|}\left(\frac{Z^*}{Z}\right)^2
\end{align}
to account for differences in nuclear charge screening from ionised backgrounds. Here $\lambda_D$ is the Debye length of the background ions, and $a_s=1.4 a_B Z^{-1/3}$ describes charge screening from atomic electrons where $a_B$ denotes the Bohr radius. The bremsstrahlung photon emission direction is sampled using a \texttt{Geant4} algorithm, which draws a direction according to the Tsai differential cross section.\cite{code:geant4:tsai1, code:geant4:tsai2} 

Two models for elastic scatter have been implemented: a hybrid-style approach used in previous hybrid-PIC codes, \cite{hybrid:Davies:1997, hybrid:Davies:2002} and a \texttt{Geant4}-style approach using an Urban algorithm adapted to the PIC framework. \cite{theory:Urban} The hybrid model solves the Fokker-Planck equation in the limit of low $Z$ targets and neglects large-angle scattering, deriving an expected deflection, $\Delta\theta$ over time $\Delta t$
\begin{align}
  \Delta \theta = \Gamma(t) \sqrt{\frac{Z^2e^4n_i \gamma m_e}{2 \pi \epsilon_0^2 p^3} \ln\left(\frac{4\epsilon_0 h p}{Z^{1/3}m_e e^2}\right)\Delta t}
\end{align}
where $\Gamma(t)$ is a random number drawn from a standard normal distribution. The Urban multiple scattering approach uses model functions which match the angular distribution moments of Lewis theory, \cite{theory:Lewis} and provides an empirical fit for mapping large angle scattering onto experimental results.\cite{theory:muonScat} The full Urban model modifies both angle and position at the end of each step, to account for scattering within the step. Steps within hybrid-PIC simulations are shorter as they are constrained to a single cell, so we neglect the spatial deviation in the PIC implementation.

\section{Benchmarking} \label{sec:code:benchmarking}

\begin{figure}
	\begin{center}
		\includegraphics[trim={0 0 0 0},clip,scale=0.5]{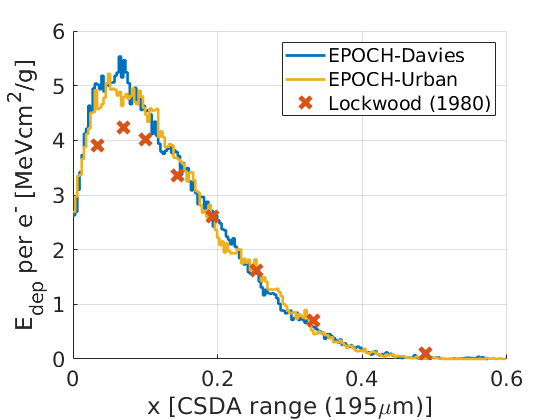}
	\end{center}
	\caption{Energy deposition per incident electron from a 0.5 MeV electron beam injected into a Ta target. Lockwood experimental data is compared to hybrid-PIC simulations running different elastic scatter algorithms.}
\label{fig:benchmark:Lockwood}
\end{figure}

The first benchmark considers the experimental results of Lockwood \textit{et al}, which measured energy deposition as a function of depth in a variety of targets. \cite{benchmark:Lockwood} In FIG. \ref{fig:benchmark:Lockwood}, the hybrid-PIC code attempts to recreate their data for a 0.5 MeV electron beam at normal incidence on a Ta target. This was performed at low electron currents which give negligible resistive fields, and bremsstrahlung and $\delta$-ray emission may also be neglected at these electron energies. This benchmark mainly tests the ionisation energy loss and elastic scatter routines. The 3D simulation window ($x\times y \times z$) spanned $150\times2\times2$ \SI{}{\micro m^3} ($256\times 8 \times 8$ cells), with open boundary conditions in $x$, and periodic boundaries in $y$ and $z$. In the first timestep, 50 macro-electrons of unit weight were injected into each cell on the $x_\text{min}$ boundary, and the simulation ran for 2 ps. The deposited grid energy was deduced from the final electron temperature distribution and the heat capacity used in the simulations, and was summed over all cells which shared an $x$ position. These simulations were performed for both Davies \cite{hybrid:Davies:2002} and Urban \cite{code:geant4:2003, code:geant4:2006, code:geant4:2016} elastic scatter models, and show a reasonable agreement with the Lockwood data. As the Davies simulation ran roughly 3 times faster, we have opted for this elastic scatter model in this paper.

The hybrid field solver, Ohmic heating, reduced Lee-More model and laser-accelerated electron injection were benchmarked against experiments with the Vulcan petawatt laser. Evans \textit{et al} obtained a temperature-depth curve with data from shots on multiple plastic targets, where the temperature was measured from a 0.2 \SI{}{\micro m} Al tracer layer buried at different depths. \cite{benchmark:Evans} This was recreated in the hybrid-PIC code using a simulation window which spanned $32.2\times 20 \times 20$ \SI{}{\micro m^3} ($322\times 40 \times 40$ cells), for a target which was Al between $x=28$ and 28.2 \SI{}{\micro m}, and plastic otherwise. The peak laser intensity was estimated to be $3.1\times 10^{20}$ \SI{}{Wcm^{-2}}, with a temporal fwhm, $t_{fwhm}=800$ fs, and a spatial radial fwhm, $r_{fwhm} = 10$ \SI{}{\micro m}. To fit the data, we assume $\eta_{l\rightarrow e} = 0.04$. To estimate the peak $T_e$, the slow thermal exchange with ions has been neglected and the final $T_e$ values are recorded. 

\begin{figure}
	\begin{center}
		\includegraphics[trim={0 0 0 0},clip,scale=0.5]{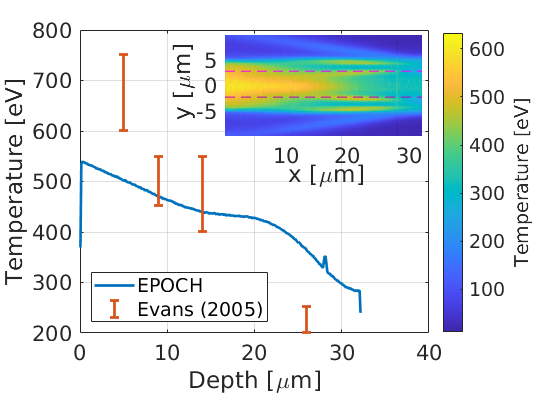}
	\end{center}
	\caption{The temperature distribution of plastic targets with Al tracer layers after exposure to Vulcan shots ($3.1\times 10^{20}$ \SI{}{Wcm^{-2}}). Experimental data is compared to the electron temperature from 3D hybrid-PIC simulations, averaged over the central $11\times 11$ cells in $y$ and $z$ for a given $x$, after 1.57 ps. A 2D heatmap of the temperature averaged over the central 11 cells in $z$ is provided in the insert, where the pink lines denote the central 11 cells in $y$. }
\label{fig:benchmark:Evans}
\end{figure}

The central $T_e(x)$ values are plotted in FIG. \ref{fig:benchmark:Evans}, and show a reasonable fit to the experimental results. The presence of the Al tracer layer at 28 \SI{}{\micro m} demonstrates the complex target capability of the code, and shows only a small increase in the temperature at this point. Deviations from experiment are attributed to the rough approximations in the injected electron characteristics, as real injections will be complicated by pre-plasmas and imperfect focal spots.

The final benchmark attempts to recreate the experimental bremsstrahlung photon number spectrum into a 40$\degree$ forward cone, from $4\times 10^{20}$ \SI{}{Wcm^{-2}} Vulcan shots on thick Au targets. \cite{benchmark:Clarke} The code modelled a 3 mm $\times 100^2$ \SI{}{\micro m^2} Au solid (cubic cells of length 0.7 \SI{}{\micro m}), and ran to 1.2 ps. Hot electrons were injected with $t_{fwhm}=800$ fs, $r_{fwhm}=5$ \SI{}{\micro m}, and $\eta_{l\rightarrow e}$ = 0.3. 

\begin{figure}
	\begin{center}
		\includegraphics[trim={0 0 0 0},clip,scale=0.5]{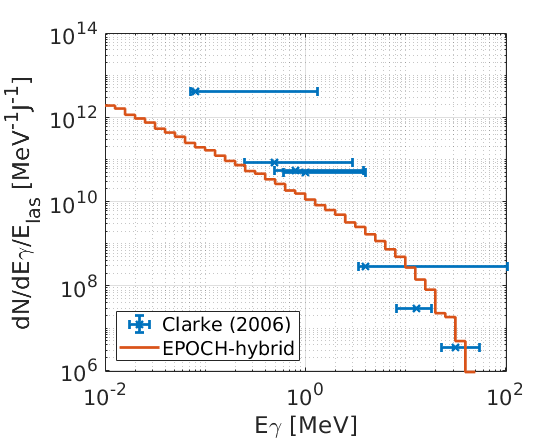}
	\end{center}
	\caption{Number spectrum of X-ray photons from a $4\times 10^{20}$ \SI{}{Wcm^{-2}} shot on a 3 mm Au target, for X-rays falling within a $40\degree$ cone ($20\degree$ half-angle) about the injection direction. Experimental data is compared to an equivalent run using the hybrid-PIC code.}
\label{fig:benchmark:Clarke}
\end{figure}

FIG. \ref{fig:benchmark:Clarke} shows the number spectrum of bremsstrahlung photons created with angle less than 20$\degree$ to the mean injection direction. While we expect to over-estimate the low energy bremsstrahlung emission as our code lacks photoelectric attenuation, \cite{theory:Photoelectric} we see that low energy X-rays are actually under-estimated here. When looking at the bremsstrahlung emission from electron beams on uniform targets, the hybrid code matches equivalent runs in \texttt{Geant4}, which suggests a correct implementation. Hence, the FIG. \ref{fig:benchmark:Clarke} discrepancy is again attributed to the over-simplified electron injection model.

\newpage

\end{document}